\documentclass[prb,twocolumn]{revtex4}

\usepackage{latexsym,amsmath,graphics,graphicx}

\begin{document}
\title{The effect of the spin-orbit interaction on the band gap of half-metals}
\author{Ph.~Mavropoulos, K.~Sato, R.~Zeller, and P.~H.~Dederichs}
\affiliation{Institut f\"ur
Festk\"orperforschung, Forschungszentrum J\"ulich, D-52425 J\"ulich,
Germany}
\author{V.~Popescu and H.~Ebert} 
\affiliation{
Department Chemie/Physikalische Chemie, University of Munich,
Butenandtstr. 5-13, D-81377 Munich, Germany} 
\date{\today}

\begin{abstract}
  The spin-orbit interaction can cause a nonvanishing density of
  states (DOS) within the minority-spin band gap of half-metals around
  the Fermi level. We examine the magnitude of the effect in Heusler
  alloys, zinc-blende half metals and diluted magnetic semiconductors,
  using first-principles calculations. We find that the ratio of
  spin-down to spin-up DOS at the Fermi level can range from below 1\%
  (e.g. 0.5\% for NiMnSb) over several percents (4.2\% for (Ga,Mn)As)
  to 13\% for MnBi.
\end{abstract}

\maketitle

Half-metals are materials showing spin ferromagnetism, but with the
exotic property of presenting a metallic density of states (DOS) only
for the one spin direction (usually majority), in parallel with a
clear band gap around the Fermi level $E_F$ for the other
direction. Experimentally, the first of the kind to be reported were
half-metallic Heusler alloys by de Groot {\it et al.}~in
1983.\cite{deGroot83} More recently, other half-metals were found,
such as CrO$_2$,\cite{Soulen98}
La$_{0.7}$Sr$_{0.3}$MnO$_3$,\cite{Soulen98} or diluted magnetic
semiconductors.\cite{Ohno98} Furthermore, ordered zinc-blende CrAs and
CrSb were fabricated by molecular beam epitaxy\cite{Akinaga00} and
relevant calculations suggest these and many similar zinc-blende
systems to be half-metallic.\cite{Shirai00,Galanakis03} So has been
the case with a variety of Heusler alloys.\cite{Galanakis02} The
rising interest in half-metallic systems is partly caused by their
potential applications in spin electronics,\cite{Wolf01} since
junctions made of half-metals should theoretically present 100\%
magnetoresistance or 100\% spin polarized currents.

Although the {\it ab-initio} calculations show the presence of
half-metallicity in this variety of materials, with a clear band gap
around $E_F$ for the minority spin, this can be partly suppressed by,
{\it e.g.}, defects,\cite{Ebert91} spin excitations at increased
temperature, or non-quasiparticle states.\cite{Irkhin94} Even at low
temperature and in the defect-free case, there is still the question
of spin-orbit interaction. Although this is weak and in many cases can
be neglected, in principle it should couple the two spin channels so
that the presence of states around $E_F$ for the majority spin is
partly reflected in the minority DOS. Thus one cannot have a real gap,
but a small DOS, its smallness depending on the strength of the
spin-orbit coupling for each material; this effect could have also
implications in the ideas about spin-dependent transport and
spintronics devices.  It is the purpose of this paper to give a
quantitative picture for this effect for a few typical systems, and to
argue if and how crucially this could affect spin-dependent transport.
As model systems we have chosen the Heusler alloy NiMnSb, the ordered
zinc-blende alloys CrAs, CrSb and MnBi, and the chemical disorder in
diluted magnetic semiconductor Ga$_{0.95}$Mn$_{0.05}$As.

Our first-principles calculations are based on the screened
Korringa-Kohn-Rostoker (KKR) Green function method, incorporating
fully relativistic effects (from where the spin-orbit interaction
emerges). For the description of the diluted magnetic semiconductors,
the coherent potential approximation (CPA) was also employed. Details
about the method can be found in Ref.~\onlinecite{Papanikolaou02}. In
the calculations we have used the atomic sphere approximation (ASA),
in which it is assumed that the potential around each atomic site is
spherically symmetric. In the open structures, such as the
zinc-blende, the large empty interstitial space is accounted for by
placing at appropriate positions empty spheres (treated as atomic
spheres with no nuclear charge). In the expansion of the Green
function in local orbitals a cutoff of $l_{\mathrm{max}}=3$ was taken
for the ordered compounds and of $l_{\mathrm{max}}=2$ for the CPA
calculations. We have employed both the fully relativistic (solving
the Dirac equation) and the scalar relativistic\cite{Koelling77}
methods for comparison. The latter includes all other relativistic
effects except spin-orbit coupling, so that the two spin channels are
decoupled and half-metals appear with a clean gap for spin-down around
$E_F$.

The DOS $n_{\sigma}(E)$ per spin $\sigma$ is related to the retarded Green function
$G_{\sigma}(\vec{r},\vec{r'};E)$ via
\begin{equation}
n_{\sigma}(E) 
= -\frac{1}{\pi}\mathrm{Im}\int G_{\sigma}(\vec{r},\vec{r};E)\,d^3r
\label{eq:0.5}
\end{equation}
Usually for the calculation a very small imaginary part $i\epsilon$ is
added to the real energy $E$ and the Green function and DOS are
evaluated at $E+i\epsilon$.  In this way the DOS of every eigenstate
is no more a delta function of the energy but is rather smoothed via
a Lorentzian broadening. This method has the advantage of needing
fewer $\vec{k}$ points for the integration in the Brillouin zone to
obtain a smooth DOS, saving computational time. Nevertheless there is
the disadvantage that, in the presence of a band gap, the band edges
are not sharp and one also obtains a very low nonzero DOS within the
gap due to the Lorentzian broadening.  In our case this causes a
problem, since the DOS within the minority gap due to spin-orbit
coupling is also small and drowned in the background of the Lorentzian
broadening. To overcome this difficulty one would have to go much
closer to the real energy axis, which would demand too many $\vec{k}$
points. We have used an alternative, which saves calculation time:
Since the Green function is analytical in the complex energy plane
$E_z$, we are able to extrapolate to the real energy axis by knowing
the variation of the Green function for $E_z$ along a line parallel to
the real axis. To our knowledge, this method was first proposed by
Gray and Kaplan,\cite{Gray86} who also report on the numerical
limitations. The resulting accuracy within the gap is much better than
that of the traditional method, and good enough for our purposes,
although sometimes numerical fluctuations of the DOS are visible.

In our calculations we solve the Dirac equation, thus treating the
spin-orbit term exactly. But for a better understanding we give now a
brief description of the first order effect of the interaction in a
Schr\"odinger picture. We remind that the spin-orbit coupling of the
two spin channels is related to the unperturbed potential $V(r)$
around each atom via the angular momentum operator $\vec{L}$ and the
Pauli spin matrix $\vec{\sigma}$:
\begin{equation}
V_{\mathrm{so}}(r)=\frac{1}{2m^2c^2}\frac{\hbar}{2}
\frac{1}{r}\frac{dV}{dr}\,\vec{L}\cdot\vec{\sigma}
= \left( 
\begin{tabular}{cc}
$V_{\mathrm{so}}^{\uparrow\uparrow}$   & $V_{\mathrm{so}}^{\uparrow\downarrow}$ \\
$V_{\mathrm{so}}^{\downarrow\uparrow}$ & $V_{\mathrm{so}}^{\downarrow\downarrow}$
\end{tabular}
\right)
\label{eq:1.0}
\end{equation}
Analyzed in spinor basis, it takes the $2\times 2$ matrix form above.
We denote the two spin directions with $\uparrow$ and $\downarrow$,
the unperturbed crystal hamiltonian for the two spin directions as
$H^{0\uparrow}$ and $H^{0\downarrow}$, and the unperturbed Bloch
eigenfunctions as $\Psi_{n\vec{k}}^{0\uparrow}$ and
$\Psi_{n\vec{k}}^{0\downarrow}$. Then the Schr\"odinger equation for
the perturbed wavefunction
$\Psi_{n\vec{k}}=(\Psi_{n\vec{k}}^{\uparrow},\Psi_{n\vec{k}}^{\downarrow})$
reads
\begin{eqnarray}
\left(
\begin{tabular}{cc}
$H^{0\uparrow}+V_{\mathrm{so}}^{\uparrow\uparrow} -E$
& $V_{\mathrm{so}}^{\uparrow\downarrow}$ \\
$V_{\mathrm{so}}^{\downarrow\uparrow}$  &
$H^{0\downarrow}+V_{\mathrm{so}}^{\downarrow\downarrow} -E$
\end{tabular}
\right)
\left(
\begin{tabular}{c}
$\Psi_{n\vec{k}}^{\uparrow}$ \label{eq:2.0}\\
$\Psi_{n\vec{k}}^{\downarrow}$
\end{tabular}
\right)
=0
\label{eq:2.1}
\end{eqnarray}
The potential terms $V_{\mathrm{so}}^{\uparrow\downarrow}$ and
$V_{\mathrm{so}}^{\downarrow\uparrow}$ are responsible for flipping
the spin. Within the gap region of the half-metal, where no spin-down
states of the unperturbed hamiltonian exist, the spin-down solution
can be found in first order by solving Eq.~(\ref{eq:2.1}) for
$\Psi_{n\vec{k}}^{(1)\downarrow}$ (the index (1) stands for the
first-order solution). The solution reads formally:
\begin{equation}
\Psi_{n\vec{k}}^{(1)\downarrow} =
-\frac{1}{H^{\downarrow}+V_{\mathrm{so}}^{\downarrow\downarrow} - E_{n\vec{k}}^{0\uparrow}} 
V_{\mathrm{so}}^{\downarrow\uparrow} \Psi_{n\vec{k}}^{0\uparrow}
\label{eq:3.0}
\end{equation}
with $E_{n\vec{k}}^{0\uparrow}$ the eigenenergy of the state
$\Psi_{n\vec{k}}^{0\uparrow}$.  We see that in the gap region the
spin-down intensity is a weak image of the band structure
$E_{n\vec{k}}^{0\uparrow}$ of the spin-up band. Since the DOS is
related to $|\Psi_{n\vec{k}}^{(1)}|^2$, it is expected that within the
gap the DOS has a quadratic dependence on the spin-orbit coupling
strength: $n_{\downarrow}(E) \sim
(V_{\mathrm{so}}^{\downarrow\uparrow})^2$. But there is also a
modification by the term
$-1/(H^{\downarrow}+V_{\mathrm{so}}^{\downarrow\downarrow}
-E_{n\vec{k}}^{0\uparrow})$ (this is actually the Green function),
which increases the weight of the states close to those $\vec{k}$
points where the unperturbed spin-up and spin-down bands cross, {\it
  i.e.}, $E^{0\uparrow}_{n\vec{k}} = E^{0\downarrow}_{n'\vec{k}}$. To
see this we can rewrite Eq.~(\ref{eq:3.0}) by expanding the Green
function in spectral representation:
\begin{eqnarray}
\Psi_{n\vec{k}}^{(1)\downarrow}(\vec{r}) &=& 
\int d^3r' \sum_{n'}
\frac{\Psi_{n'\vec{k}}^{(0)\downarrow}(\vec{r})
\Psi_{n'\vec{k}}^{(0)\downarrow *}(\vec{r'})}
{E_{n\vec{k}}^{0\uparrow}-E_{n'\vec{k}}^{0\downarrow}} \,
V_{\mathrm{so}}^{\downarrow\uparrow}(\vec{r'})
\Psi_{n\vec{k}}^{(0)\uparrow}(\vec{r'}) \nonumber\\
&=&
\sum_{n'} 
\frac{\langle\Psi_{n'\vec{k}}^{0\downarrow}|
V_{\mathrm{so}}^{\downarrow\uparrow}|\Psi_{n\vec{k}}^{0\uparrow}\rangle}
{E_{n\vec{k}}^{0\uparrow}-E_{n'\vec{k}}^{0\downarrow}}
\Psi_{n'\vec{k}}^{0\downarrow}(\vec{r})
\label{eq:4.0}
\end{eqnarray}
Here, the summation runs only over the band index $n'$ and not over
the Bloch vectors $\vec{k}'$, because Bloch functions with
$\vec{k}'\neq\vec{k}$ are mutually orthogonal. Close to the crossing
point $E_{n\vec{k}}^{0\uparrow}-E_{n'\vec{k}}^{0\downarrow}$ the
denominator becomes small and the bands strongly couple. Then one
should also consider higher orders in the perturbation
expansion. Since at the gap edges there exist spin-down bands of the
unperturbed hamiltonian, this effect can become important near the gap
edges.

\begin{figure}[t]
\begin{center}
\includegraphics[angle=90,width=6.9cm]{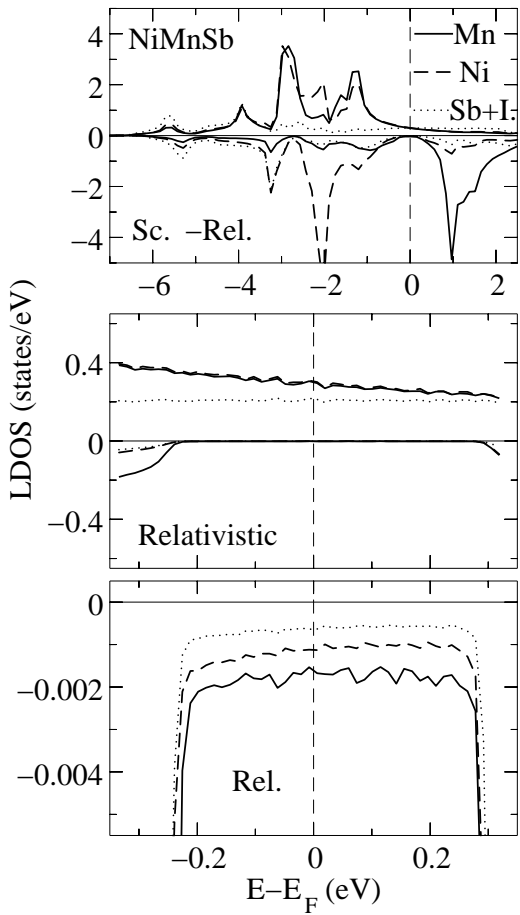}
\caption{DOS of NiMnSb. Top: scalar relativistic (I. refers to
  interstitial volume of the empty sphere). Middle: fully
  relativistic, focused in the minority-gap region around $E_F$. In
  this scale, the spin-orbit effect is not visible. Bottom: fully
  relativistic, focused on the minority states within the ``gap''. The
  fluctuations reflect the numerical accuracy of the method (see
  text). Around $E_F$, the spin-down DOS is of the order of 0.5\% of
  the spin-up DOS. Close to the ``gap'' edges, the spin-orbit induced
  minority DOS increases strongly as explained in the text. Negative
  numbers in the DOS axis correspind to minority spin.}
\label{fig:1}
\end{center}
\end{figure}
Now we proceed to the analysis of our results. The results for the
Heusler alloy NiMnSb, which was found by de Groot {\it et al.}~in 1983
to be half-metallic,\cite{deGroot83} are shown in Fig.~\ref{fig:1}. In
the top panel, the atom-resolved DOS is shown, following a scalar
relativistic calculation. The Fermi level is within the band gap for
minority spin and the material is half-metallic. In this scale, the
relativistic calculation shows no difference in the DOS compared to
the scalar relativistic. In the middle panel the gap region is shown
in a relativistic calculation.  Again, no minority DOS is seen in this
scale. But when we magnify the minority DOS in the lower panel, it
becomes clear that the DOS is nonzero in the gap region --- it is of
the order of $0.5\%$ of the majority DOS in the vicinity of $E_F$. We
have checked the DOS in this scale for the scalar relativistic
calculation and have found it to be zero, thus we conclude that we see
here a spin-orbit effect and not an artificial Lorentzian broadening
due to the imaginary part of the energy in the calculation. The
fluctuations of the Mn DOS (full line) are due to numerical noise of
the $\vec{k}$ integration and extrapolation scheme, and mark its limit
of accuracy.

\begin{figure}[t]
\begin{center}
\includegraphics[angle=90,width=6.9cm]{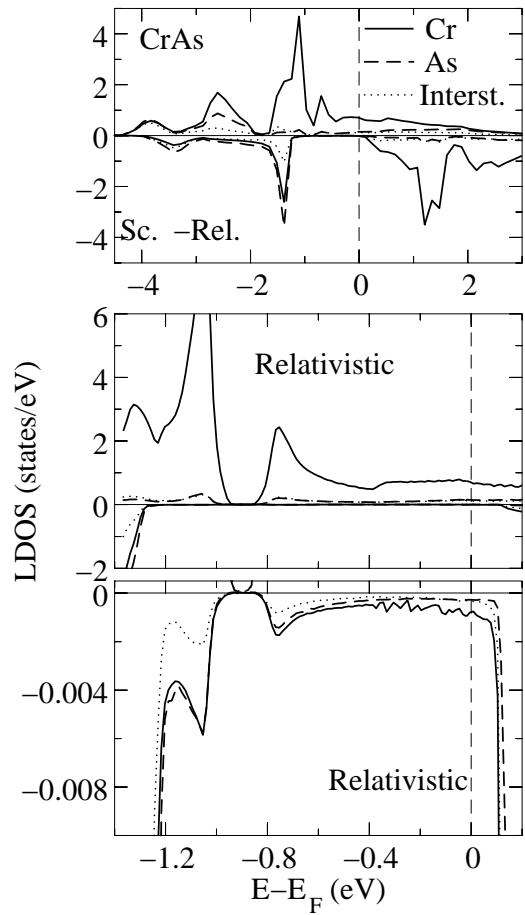}
\caption{DOS of zinc-blende CrAs. Top: scalar relativistic. Middle: fully
  relativistic, focused in the minority-gap region around $E_F$. In
  this scale, the spin-orbit effect is not visible. Bottom: fully
  relativistic, focused on the minority states within the ``gap''. The
  reflection of spin-up into spin-down DOS is clear at the peaks at
  $-1\ $eV and $-0.8\ $eV. In the middle of the gap, the spin-down
  DOS is of the order of 0.2\% of the spin-up DOS. Negative
  numbers in the DOS axis correspind to minority spin.}
\label{fig:2}
\end{center}
\end{figure}
The next example is the ordered half-metallic alloy CrAs in the
zinc-blende structure. This was invented by Akinaga and
co-workers\cite{Akinaga00} in 2000 and found to be half-metallic. Our
results are presented in Fig.~\ref{fig:2}. In the top panel, the
scalar relativistic atom-resolved DOS is shown. The Fermi level is
within the minority-spin gap, close to the conduction band edge. (The
latter is mainly of $d$ character, containing the wavefunctions of the
$e_g$ subspace~\cite{Galanakis03}). Focusing in the range of the gap,
we can see the DOS in more detail in the middle panel in a
relativistic calculation. Below $-1\ \mathrm{eV}$ one recognises the
majority $d$ states of $e_g$ character.\cite{Galanakis03} Around
$-0.9\ \mathrm{eV}$ the DOS has a minimum, as the $e_g$ states end and
the $d$ states of $t_{2g}$ character start, forming wide bands. In
this scale no spin-orbit effect is visible in the minority gap. But if
we focus further in energy (lower panel) it is clear that the
structure of the majority DOS is reflected in the minority DOS: one
can recognize the peak of the $e_g$ states below $-1\ \mathrm{eV}$,
then the minimum and finally the presence of the $t_{2g}$ band above
$-0.8\ \mathrm{eV}$. This is the effect of the term
$V_{\mathrm{so}}^{\downarrow\uparrow} \Psi_{n\vec{k}}^{\uparrow}$ in
Eq.~(\ref{eq:3.0}). But we also observe the rapid increase of the DOS
as the energy approaches the band edges; this comes from the term
$-1/(H^{\downarrow}+V_{\mathrm{so}}^{\downarrow\downarrow}
-E_{n\vec{k}}^{0\uparrow})$. The minority DOS in the middle of the gap
is again small, giving a minority/majority DOS ratio of $0.2\%$.  In a
similar calculation for zinc-blende CrSb (first fabricated by Zhao
{\it et al.}\cite{Zhao01}) we find that the minority DOS in the middle
of the gap is around $0.7\%$ of its majority counterpart. This rise is
expected, as the spin-orbit interaction is stronger for the $5p$
states of Sb than for the $4p$ of As. (The trend can be seen for
example in a systematic study of the spin-orbit cross section of $sp$
impurities in Mg in Ref.~\onlinecite{Papanikolaou92}, where Sb is
found to have four times higher spin-orbit cross section than As.
Since the spin-orbit interaction is in first approximation an atomic
property, the trend is similar here.)

\begin{figure}[t]
\begin{center}
\includegraphics[angle=90,width=6.9cm]{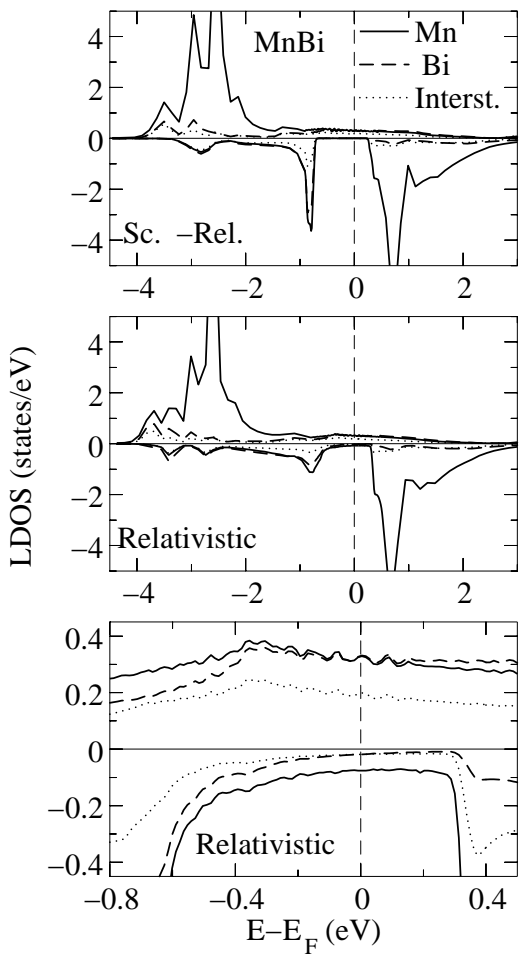}
\caption{DOS of zinc-blende MnBi. Top: scalar relativistic. Middle: Fully
  relativistic. Bottom: fully relativistic, focused on the states
  within the minority ``gap''. At $E_F$, the spin-down DOS is of the
  order of 13\% of the spin-up DOS. This higher value is expected
  since the $6p$ valence states of Bi result in a strong spin-orbit
  effect. Negative numbers in the DOS axis correspind to minority
  spin.}
\label{fig:3}
\end{center}
\end{figure}
A recent theoretical investigation\cite{Xu02} has suggested the
ordered zinc-blende alloy MnBi as a good candidate for
half-metallicity. Although this has not yet been fabricated, we take
it as a good example to show the spin-orbit effect, since Bi is a
heavy element with strong spin-orbit interaction because of the 6p
states. The scalar and fully relativistic DOS are shown in
Fig.~\ref{fig:3}. It is evident that here the spin-orbit interaction
has a much stronger influence. The minority DOS at $E_F$ is of the
order of $13\%$ of its majority counterpart, a huge quantity in
comparison to the analogous percentage in CrAs or NiMnSb.  The
relativistic effect shows up also in the orbital magnetic moment,
being $0.11\mu_B$ and mostly concentrated at the Mn atom, while the
spin moment remains at $4.0\mu_B$ per unit cell as in the scalar and
non-relativistic case. We note that, although the spin-orbit
interaction is strong here, the majority spin is still dominating at
$E_F$. Thus MnBi would be a useful material for spin electronics
despite the spin-orbit effect, as its spin polarization at $E_F$ is
robust against moderate volume change as pointed out in
Ref.~\onlinecite{Xu02}.

As a final example we consider the diluted magnetic semiconductor
(DMS) Ga$_{0.95}$Mn$_{0.05}$As. This material, together with other
similar DMS, has been extensively studied in recent years, because it
presents ferromagnetism at such low Mn concentrations (discovered by
Ohno\cite{Ohno98} in 1998). Ideally, the Mn impurities substitute Ga
atoms in the GaAs lattice. Then the Mn $3d$ states of the $t_{2g}$
representation ($d_{xy}$, $d_{yz}$, and $d_{xz}$) hybridize with the
$p$ states of the As neighbors, forming bonding and antibonding
$p$-$d$ hybrids and from them states within the GaAs gap for the
spin-up direction. Thus the alloy becomes half-metallic with $E_F$
close to the valence band edge.  Although there is no consensus yet on
the physical origin of the ferromagnetic ordering, which seems to
depend on the energetic position of the Mn $d$ states,\cite{footnote}
we have proceeded using the LDA result. We find that the spin-down DOS
at $E_F$ is $4.2\%$ of its spin-up counterpart if we include
spin-orbit effects, while the system is half-metallic in the
nonrelativistic solution. To understand this relatively high
percentage of spin-down states, we observe that $E_F$ is very close to
the valence band edge. There, the $p$ states of As give an important
spin-orbit effect. This is known to split the valence band edge of
GaAs in two subspaces, one with total (spin + orbital) angular
momentum $j=3/2$ and one with $j=1/2$, so that the valence band edge
has no pure spin character.\cite{Yu} Similar calculations on Mn-doped
ferromagnetic zinc-blende GaN gave us no spin-orbit effect at $E_F$
within numerical accuracy.  The difference to (Ga,Mn)As is that the
$2p$ valence states of N compared to the $4p$ of As (i) have a much
lower spin-orbit interaction, and (ii) are lower in energy so that the
Mn impurity $d$ band is fully within the band gap, away from the
N-dominated valence band states, and the spin-orbit effect weakens.

In summary, we have performed first-principles calculations in order
to investigate the effect of the spin-orbit coupling to the spin-down
band gap of half-metallic systems. As typical half-metallic systems,
describable by the LDA, we have chosen the Heusler alloy NiMnSb, the
ordered zinc-blende alloys CrAs, CrSb, and MnBi, and the diluted
magnetic semiconductors (Ga,Mn)As and (Ga,Mn)N. We find that the
majority-spin states are partly reflected into the minority band gap.
The intensity of the DOS for minority electrons in this energy region
depends mainly on the strength of the spin-orbit coupling, heavier
$sp$ elements resulting in a higher ratio of minority/majority spin
DOS. Thus we see a trend of increasing min./maj. DOS ratio in the
middle of the gap as we go from CrAs (0.2\%) to CrSb (0.7\%) and
finally to MnBi (13\%). In NiMnSb we find a ratio of 0.2\%. Also, it
is important how deep the Fermi level lies within the gap, since
majority states close to the gap edges are more drastically
spin-flipped than states deep in the gap.  This results in high
minority-spin DOS (4.2\% ratio) for (Ga,Mn)As where $E_F$ is very
close to the valence band edge. We have also explained these effects
within a first-order approximation. For transport applications in
spintronics we conclude that even in compounds with strong spin-orbit
effect (MnBi or (Ga,Mn)As) the minority DOS at $E_F$ is still
dominated by its majority counterpart (even in MnBi the ratio is only
13\%), whence current should be transported practically only by
majority states.  Thus in applications other effects, such as gap
states due to impurities,\cite{Ebert91} stalking faults, or
interfaces, could be more important to eliminate by sample
improvements.

\begin{acknowledgments}
  This work was supported by the RT Network of {\em Computational
    Magnetoelectronics} (contract RTN1-1999-00145) of the European
  Commission.  One of us (VP) gratefully acknowledges the financial
  support of the Deutsche Forschungsgemeinschaft withing the
  DFG-F\"orderprojekt FOG 370/2-1
  ''Ferromagnet-Halbleiter-Nanostrukturen: Transport, elektrische und
  magnetische Eigenschaften''.
\end{acknowledgments}


\end{document}